\documentclass[a4paper,11pt]{article}
\usepackage{pos}
\usepackage{amsmath}
\usepackage{caption}
\usepackage{float}
\usepackage{wrapfig}

\title{End-to-End Tests of the Sensitivity of IceCube to the Neutrino Burst from a Core-Collapse Supernova}
\ShortTitle{IceCube Supernova Data Challenges}

\author{The IceCube Collaboration \\{\normalsize \normalfont(a complete list of authors can be found at the end of the proceedings)}}
\emailAdd{sgriswold@icecube.wisc.edu}

\abstract{The next galactic supernova presents a once-in-a-lifetime opportunity to obtain detailed information about the explosion of a star and the extreme conditions found within its core. A core-collapse supernova will produce a neutrino burst visible up to half a day before the electromagnetic radiation from the explosion, so the burst will provide an early warning for optical follow-up. Since local supernovae are exceedingly rare, it is critical that neutrino detectors provide prompt alerts after the arrival of a burst. The IceCube Neutrino Observatory operates with $>$99\% uptime and is sensitive to a variety of supernova models at levels >10$\sigma$ within the Milky Way. IceCube will issue supernova alerts in real time. IceCube’s high sensitivity to supernovae, near perfect uptime, and ability to issue prompt alerts makes it a critical component of the worldwide network of detectors known as the SuperNova Early Warning System (SNEWS 2.0). A ``Fire Drill'' system was designed to inject simulated supernova signals into the IceCube online system. We will discuss IceCube’s sensitivity to supernovae near the Milky Way, and describe the data challenges used to ensure the readiness of IceCube and its operators. We will also discuss the coordination of IceCube alerts and data challenges with SNEWS 2.0.

\vspace{4mm}
\textbf{Corresponding authors:}
Spencer Griswold$^{1*}$\\
{$^{1}$ \itshape Dept. of Physics and Astronomy, University of Rochester\\
  206 Bausch \& Lomb Hall P.O. Box 270171, Rochester, NY 14627, USA}\\[4mm]
$^*$ Presenter

\FullConference{37$^{\rm{th}}$ International Cosmic Ray Conference (ICRC 2021)\\
		July 12th -- 23rd, 2021\\
		Online -- Berlin, Germany}
}

\begin{document}
\maketitle

\section{Introduction}
A galactic core collapse supernova (CCSN) will produce a high intensity burst of all flavors of neutrinos. Neutrinos produced by a CCSN may be used to probe the core structure and core equation of state of the exploding star. It will also produce insight into fundamental neutrino physics and potentially physics beyond the Standard Model \cite{Horiuchi:2017sku}. The neutrino burst from a CCSN is expected to predate the arrival of photons from the explosion by hours to days, enabling neutrinos to serve as an early indicator for optical astronomers. An early warning is crucial to ensure astronomers perform the first observation of the onset of the supernova explosion, and in particular capture the breakout burst \cite{Nakamura:2016kkl}. The neutrino burst may also arrive in tandem with gravitational waves (see \cite{Nakamura:2016kkl}), enabling measurement of the absolute mass of the neutrino \cite{Vissani:2010fg}. The next galactic supernova presents a once-in-a-lifetime opportunity to make a groundbreaking multi-messenger astrophysical measurement. Because these events are exceedingly rare, it is necessary to prepare for their arrival.

IceCube instruments a cubic kilometer of ice at the geographic South Pole using a lattice of 5,160 digital optical modules (DOMs). The DOMs are deployed in the Antarctic glacier at depths between 1.5~km to 2.5~km beneath the surface. They are arranged into 86 strings spaced 125~m apart, with 60 DOMs per string spaced 17~m vertically. Each DOM uses a 10" photomultiplier tube (PMT) to capture Cherenkov photons produced by particle interactions in the ice. A sub-set of 8 strings are equipped with DOMs with 35\% higher quantum efficiency. This sub-array, DeepCore, is deployed in the center of the detector with average inter-string spacing of 72~m and inter-DOM spacing ranging from 7~m to 10~m \cite{Aartsen:2016}. The dense arrangement and efficiency of DeepCore increases IceCube's sensitivity to lower energy neutrinos \cite{Collaboration:2011ym}. In the case of a CCSN, the neutrino burst will be comprised of $\mathcal{O}(10\text{ MeV})$ neutrinos and the dominant interaction in the ice will be inverse beta decay (IBD) of $\bar{\nu}_e$ events. At these energies, roughly 1 out of 400 IBD Cherenkov photons produced in the ice will be registered by the DOMs. This signal is not bright enough to trigger IceCube's simple multiplicity trigger and would not produce a significant detection in any single DOM. However, the neutrino burst would affect all of the detector's instrumented volume, forming a detectable collective increase in the hit rates across all DOMs. IceCube is sensitive to the neutrino signal from a galactic CCSN at a significance level $\gg10\sigma$ \cite{Abbasi:2011ss}. 

IceCube is particularly well-suited to monitor the Milky Way for CCSNe. Since 2015, IceCube's supernova data acquisition system (SNDAQ) has operated with $>$99\% trigger-capable uptime per annum, and is supplemented with the HitSpool data buffering system \cite{Heereman:2015mbs}. SNDAQ issues alerts on supernova candidate triggers in real time, and in parallel, issues a request to HitSpool to buffer the DOM waveforms for a 90~s window surrounding the burst time \cite{Heereman:2015mbs}. Data from the HitSpool system is available for offline analysis and provides the opportunity to make measurements of the neutrino lightcurve with sub-ns precision.
%IceCube's high uptime and capability to issue prompt notification of a statistically significant CCSN trigger make the observatory a keystone of galactic supernova detection. 

IceCube is a key component of the SuperNova Early Warning System (SNEWS 2.0) \cite{Kharusi:2020ovw}, a network of neutrino detectors designed to give advanced notice of imminent photons from a nearby CCSN. By identifying coincidences between the arrival times of neutrino bursts in detectors around the world, SNEWS is intended to facilitate robust optical follow-ups of CCSNe. IceCube has coordinated its supernova alerts with SNEWS since 2009. Currently, IceCube issues alerts to SNEWS at a rate of about once per month. Since 2018, IceCube has transmitted a second ``diagnostic'' data stream of low-significance alerts to SNEWS at the rate of about 5 triggers per day. The diagnostic channel tests the latency and health of the connection between IceCube and SNEWS, and is used to test the SNEWS multi-detector coincidence software. 

In this contribution, we discuss the detection of CCSNe neutrinos at the IceCube Neutrino Observatory; the method used to test the formation of supernova triggers; and tests of IceCube's operational readiness to trigger on and respond to CCSN neutrino bursts. We also discuss future plans for coordinating data challenges with other neutrino detectors and optical telescopes through SNEWS 2.0.

\section{Supernova Detection}\label{sec:detection}
SNDAQ searches for a significant deviation of the collective average DOM rate from the expected rate of background hits. The background rate is 286~Hz per DOM, including an artificial deadtime used to optimize the signal-to-noise ratio \cite{Abbasi:2011ss}. A comprehensive description of SNDAQ is provided in \cite{Abbasi:2011ss, Heereman:2015mbs}. In brief, SNDAQ compares the instantaneous hit rate across the detector to the collective average DOM rate computed using a sliding 10-minute time window. The individual hit rates of the DOMs, ${R_i}$, where $i=1\ldots N_{DOM}$, are measured online with 2~ms resolution. The trigger is formed by summing over all DOMs and searching for statistically significant excesses above background by maximizing the likelihood
\begin{equation}\label{eq:llh}
\mathcal{L}(\Delta \mu) = \prod^{N_{DOM}}_{i=1} \frac{1}{\sqrt{2\pi}\left<\sigma_i\right>} \exp\left( -\frac{(R_i - (\left<R_i\right>+\varepsilon_i \cdot \Delta\mu))^2}{2\left<\sigma_i\right>^2}\right),
\end{equation}
where $N_{DOM}$ is 5160 and $\varepsilon_i$ is the relative efficiency of DOM $i$. (For standard DOMs $\bar{\varepsilon}_i=1.0$, and for the high quantum efficiency DOMs in the IceCube DeepCore infill detector \cite{Aartsen:2016}, $\bar{\varepsilon}_i=1.35$.) In eq.~\eqref{eq:llh}, $\langle\sigma_i\rangle^2$ is the variance in the hit rate of DOM $i$ estimated using the sliding 10-minute time window. The free parameter in the likelihood is $\Delta\mu$, the collective rise in the hit rate across all DOMs. Maximizing eq.~\eqref{eq:llh} yields
\begin{equation}\label{eq:maxllh}
    \Delta\mu = \sigma^{2}_{\Delta\mu} \sum^{N_{DOM}}_{i=1} \frac{\varepsilon_i (R_i - \left<R_i\right>)}{\left<\sigma_i\right>^2} \quad \text{ and}\quad \sigma^{2}_{\Delta\mu} = \left(\sum^{N_{DOM}}_{i=1} \frac{\varepsilon^2_i}{\left<\sigma_i\right>^2}\right)^{-1},
\end{equation}
where $\sigma^2_{\Delta\mu}$ is the estimated variance of the maximum likelihood value $\Delta\mu$. The significance of the collective rate increase is expressed in terms of the test statistic
\begin{equation}\label{eq:xi}
    \xi = \frac{\Delta\mu}{\sigma_{\Delta\mu}}.
\end{equation}

The test statistic $\xi$ is computed in central time windows of size 0.5~s, 1.5~s, 4~s, and 10~s. Each search has its own physics motivation: the 0.5~s bin is optimized for short neutrino bursts produced in ``failed'' supernovae associated with the formation of black holes \cite{O_Connor_2011}; 
the 1.5~s binning is optimized against intermediate-length neutrino bursts lasting 15~s \cite{Piegsa:2009zz}; 
the 4~s binning matches the expected time constant of neutrino emission during the proto-neutron star cooling phase \cite{Abbasi:2011ss}; and the 10~s binning covers the time span of neutrinos detected from SN1987A \cite{Abbasi:2011ss, Heereman:2015mbs}.

The analysis is performed in parallel using the four time windows, and the maximum $\xi$ across all searches is extracted for a particular alert and reported. The method was chosen to reduce binning effects on the reported alert. A ``changepoint'' trigger based on the Bayesian Blocks algorithm is also used to ensure the trigger is robust against binning artifacts \cite{Cross:2019jpb}.

The background rate is dominated by radioactive decays in the PMT and DOM glass enclosures \cite{Abbasi:2011ss}, but at high $\xi$ the trigger rate exhibits a seasonal dependence due to atmospheric muons which trigger large numbers of DOMs in the detector \cite{Baum:2013, Baum:2015drl}. The effect of atmospheric muons ranges from 3~Hz to 30~Hz per DOM depending on the density of the atmosphere above the South Pole \cite{Baum:2015drl}. The correlation between the muon rate and the size of the test statistic $\xi$ can be measured online and zeroed out in real time to produce a muon-corrected test statistic $\xi^\prime$ \cite{Baum:2017rty}. The use of $\xi^\prime$ eliminates the seasonal variation observed in the high-$\xi$ tail of triggers from SNDAQ.

\begin{figure}
    \centering
    \includegraphics[width=\textwidth]{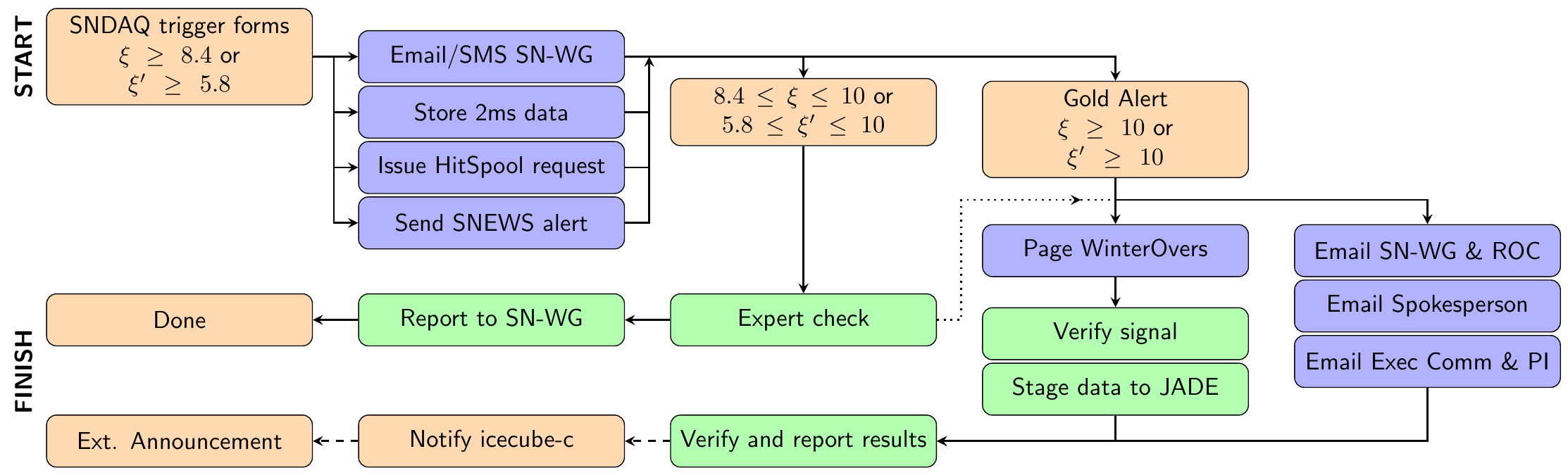}
    \caption{Simplified scheme for describing the escalation of a CCSN alert following trigger formation. Sending a SNEWS alert and issuing a HitSpool request are triggered by the same threshold condition, $\xi \geq 8.4$ or $\xi^\prime \geq 5.8$. Blue nodes represent software-based, automated steps and green nodes represent steps that require human participation.}
    \label{fig:esc_scheme}
\end{figure}

Under normal operations, $\xi$ and $\xi^\prime$ are both used to evaluate alerts and determine a course of action. This will include physically interpreting the alert's trigger, which falls to the Supernova Working group (SN-WG). The escalation scheme for handling a supernova candidate is described in Fig.~\ref{fig:esc_scheme}. When a trigger forms with $\xi\geq8.4$ or $\xi^\prime\geq5.8$, a number of automated notifications are issued, including an email to the SN-WG; a data buffering request is made through the IceCube HitSpool system; and an alert is sent to SNEWS. A ``gold'' alert has $\xi \geq 10$ or $\xi^\prime \geq 10$, and is automatically elevated to the attention of the SN-WG, the IceCube spokesperson, and the collaboration's Executive Committee. Simultaneously, on-site and off-site detector operators ensure the detector is running normally. The operators check the validity of the data and secure the data for transfer to the Northern Hemisphere via satellite link. Once the collaboration determines the supernova candidate is a valid trigger, it may decide to issue a public alert such as a GCN notice. Alerts that pass the first threshold of $\xi\geq8.4$ or $\xi^\prime\geq5.4$ but not the gold threshold may be escalated manually if deemed appropriate. The escalation scheme is designed to reduce the chance of a false positive identification and to accurately obtain physics results before issuing a public announcement.

Since the escalation scheme contains both automatic and human elements, it is important to provide regular tests to identify ``edge cases'' and other circumstances where the procedure fails. The remainder of this proceeding describes the implementation of such a test.

\section{Implementation of an offline ``Fire Drill'' Testing System}\label{sec:firedrill}
A ``Fire Drill'' system was developed to test IceCube’s supernova trigger formation and automated notification distribution. This test proceeded by performing a software injection of a simulated supernova signal into archival data which did not contain a supernova signal. The simulated hits were injected at the input of the IceCube data pipeline on the South Pole Test System (SPTS), an offline testing and development environment which runs a duplicate of the IceCube DAQ. The combined simulated and captured data was processed by the DAQ software and subsequently processed by SNDAQ, resulting in the formation of a supernova candidate trigger.

\begin{figure}
\begin{minipage}[t]{.49\textwidth}
    \centering
    \includegraphics[height=7cm]{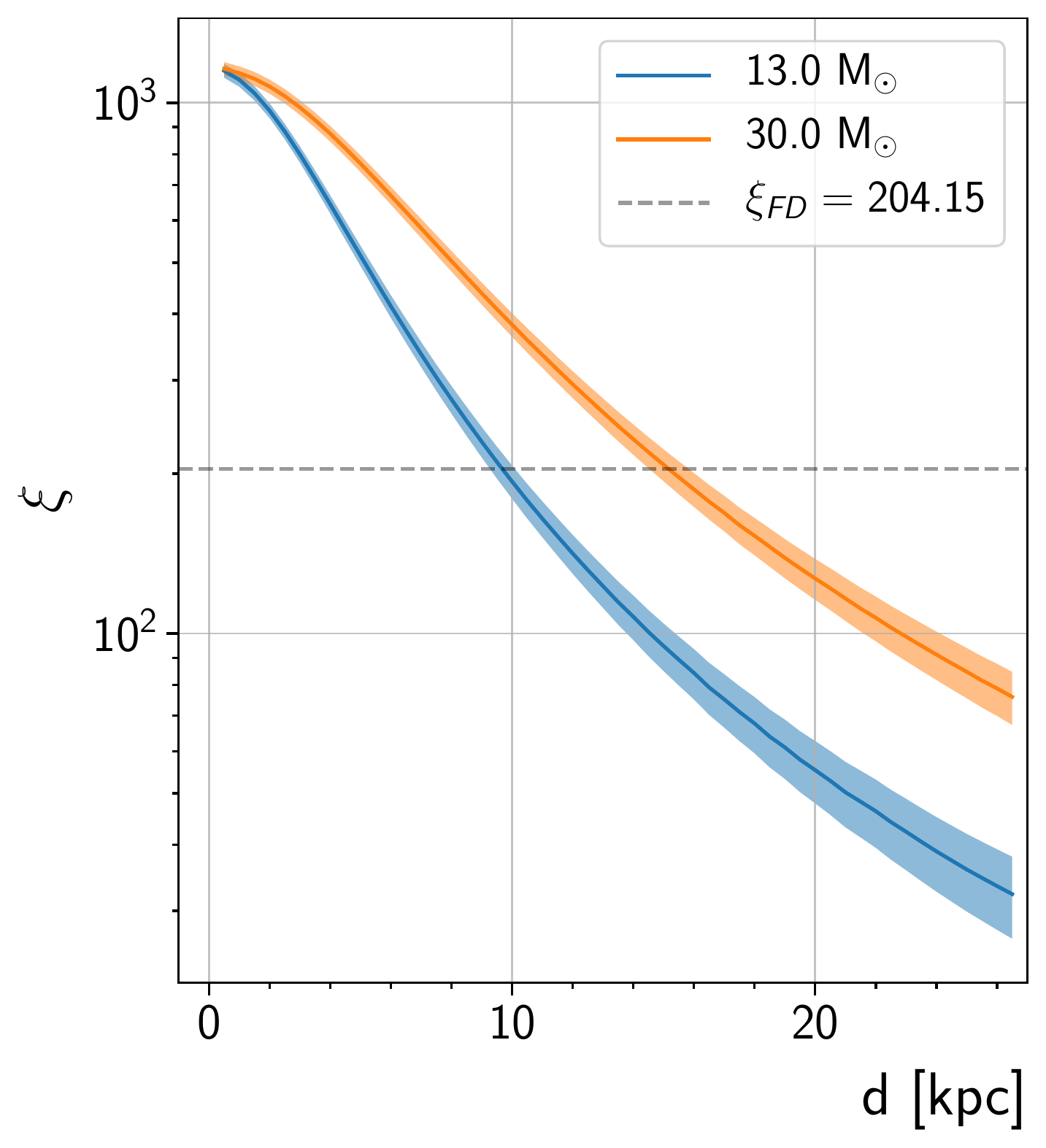}
    \captionof{figure}{Supernova triggering test statistic $\xi$ for 13 M$_\odot$ and 30 M$_\odot$ progenitors as a function of distance $d$ using models from \cite{Nakazato:2012qf}. A trigger with $\xi_\mathrm{FD}=204.15$ (dashed line) was created for the ``Fire Drill'' test.}
    \label{fig:signi}
\end{minipage}\hfill
\begin{minipage}[t]{.49\textwidth}
    \centering
    \includegraphics[height=7cm]{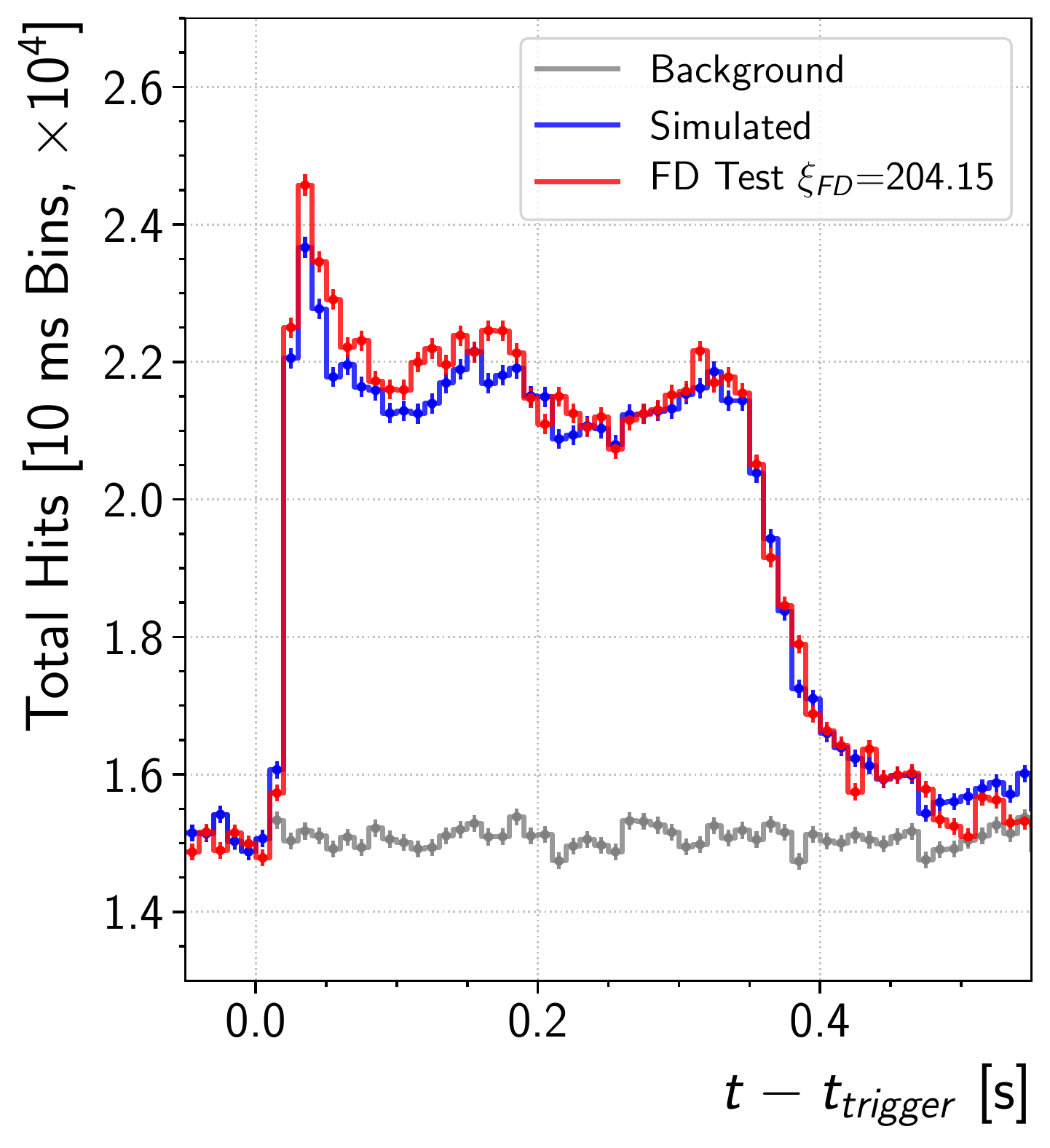}
    \captionof{figure}{Comparison of the neutrino lightcurve simulated with ASTERIA \cite{ASTERIA:2019} and the lightcurve reported by SNDAQ during the ``Fire Drill'' (FD) test.}
    \label{fig:firedrill_result}
    \end{minipage}
\end{figure}

IceCube's response to a CCSN explosion was simulated using the ASTERIA fast Monte Carlo package \cite{ASTERIA:2019} using a model from Nakazato \textit{et al.} \cite{Nakazato:2012qf} based on a 13~M$_\odot$ progenitor 10~kpc from Earth. The blue line in Fig.~\ref{fig:firedrill_result} shows the output of this simulation: the total number of hits observed across the entire detector binned in 100~$\mu$s intervals (much finer than the online resolution). This figure is discussed in detail in Sec. \ref{sec:results}. The DOMs can be treated as independent sensors, so the simulation begins by calculating the number of hits and times at which those hits occurred in each DOM. This was repeated for every DOM in the detector and lists of hit times indexed by DOM ID for each string were obtained. These lists were passed to a secondary processing script that injected the simulated signal into archival data that had been captured by HitSpool during September 2015.
\begin{figure}
    \centering
    \includegraphics[width=0.9\textwidth]{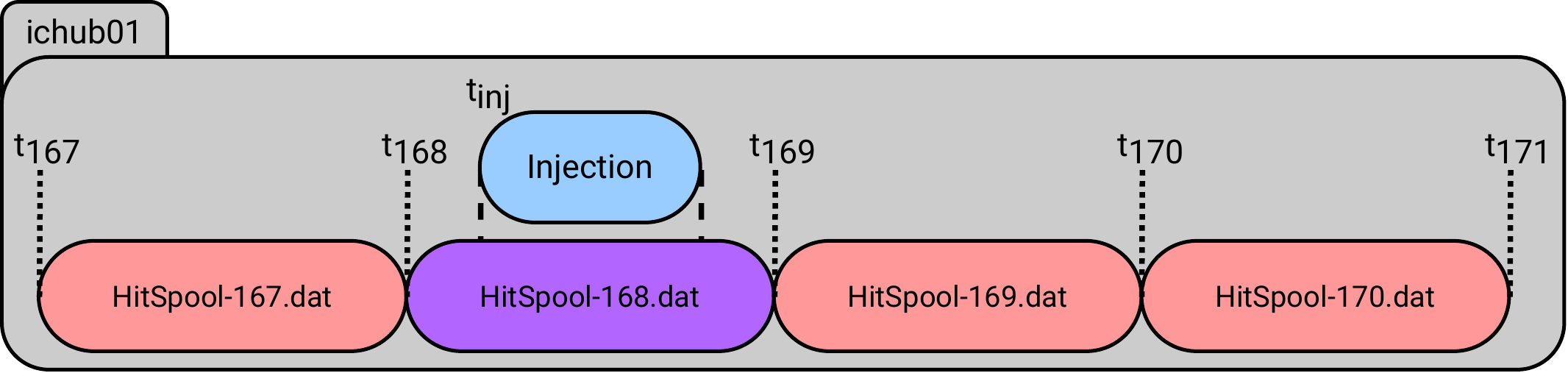}
    \caption{File selection processed use during signal injection. Each HitSpool file corresponds to 15~s worth of data while the injected signal lasts for only 10~s. Using the time at which the signal injection begins $t_{inj}$, and the start and end times of each HitSpool file, a so called 'to-be-modified' flag is constructed. For a each string, a subset of files are flagged, and later simulated hits are added to these files.}
    \label{fig:to-edit}
\end{figure}

The addition of simulated hits created new local coincidences between simulated and archival hits. Within the data, a local coincidence flag is used to indicate if a hit has occurred within 1~$\mu$s of another hit in a neighboring or next-to-neighboring DOM. To modify the archival hits, a rolling buffer was used to combine the archival and simulated data. Checks for local coincidence were made when a simulated hit was added to the front or removed from the end of this buffer. During this time, the simulated hits were reformatted from the ASTERIA output to match the HitSpool format \cite{Heereman:2015mbs}. In the case of this initial test, the simulated hits spanned a time window $\sim$10~s in duration and were added starting at 17 minutes into the 1 hour period of archival data from HitSpool. The archival data was broken into sets of files for each string of the detector, with each file representing 15~s worth of hits. To optimize the runtime of this processing step, a subset of the archival files were chosen based on their start and end time as illustrated by Fig.~\ref{fig:to-edit}. These files were processed hit-by-hit using the rolling buffer to construct a file containing the combined data. This was repeated for every string before staging the hour of data to SPTS. Using SPTS it is possible to ``replay'' the data captured by HitSpool in real time, using a so-called replay-run. During a replay-run, hits are processed by the IceCube DAQ and then other DAQ components like SNDAQ as if they were receiving a live feed of data.

\section{Results}\label{sec:results}

A replay-run using the combined data was performed on SPTS, and SNDAQ correctly formed a supernova candidate trigger in the 500~ms binning analysis with $\xi=204.15$. This is in agreement with the expectation of $\xi=193.36\pm13.91$ for a 13 M$_\odot$ progenitor at 10~kpc, shown in Fig.~\ref{fig:signi}. 
The hit rate reported by SNDAQ is shown in Fig.~\ref{fig:firedrill_result}.

This initial test also highlighted a number of low-level problems; notifications were misdirected to defunct email lists, and the automatic HitSpool request to buffer data was not properly formed. The test highlighted the need to exercise the alert escalation scheme, and next steps will focus on extending this test to the IceCube online triggering system.

\section{Future work and Coordination with SNEWS}
The first test of the signal injection ``Fire Drill'' system serves as a proof of principle for future tests and has highlighted cases where the automated components of the escalation scheme did not behave as expected. Moving forward, the ``Fire Drill'' system may be expanded to test the human elements of alert handling which require coordination with the detector operators and collaboration working groups. Future tests will include integration with IceCube's online trigger system to identify edge cases and failure states that would hinder the detection and preservation of a real supernova alert.

Ultimately, the detection of a real supernova signal will be accompanied by a physical interpretation performed by the SN-WG. Following a real alert, the SN-WG's findings would be confirmed by the experiment spokespeople and executive committee before an external announcement, such as a GCN, were issued. It is useful to perform tests with very strong signals ($\xi \gg 10$ or $\xi^\prime \gg 10$) to test the lines of communication between working groups responsible for issuing an alert. However, a real alert may present ambiguities in terms of its physical interpretation, particularly if the trigger is close to the alert threshold ($8.4 \leq \xi \leq 10$ or $5.8 \leq \xi^\prime \leq 10$). Developing a clear plan of action in this kind of ambiguous case is as important as exercising the response of the collaboration to a very clear CCSN signal. The question about whether or not to escalate such an alert to the public is a crucial one that must be carefully considered. Online testing presents the opportunity to consider a host of questions important to the validation of an alert:

\begin{itemize}
    \item Are the data of good quality? Are the DOMs experiencing a high rate of noise hits?
    \item Have the data been automatically secured by HitSpool and manually, by the detector operators?
    \item Does the supernova signal coincide with another IceCube event, like a notable cosmic ray air shower or high-energy astrophysical neutrino?
    \item How does external information, such as detection of pre-supernova neutrinos by another experiment \cite{Kharusi:2020ovw}, affect our handling of low significance alerts?
\end{itemize}

Through online testing, the answers to these questions may be approached in a way that is as realistic to the detection of a true signal as possible. With an online ``Fire Drill'' testing system in place, the opportunity then arises to perform blind tests, where the properties of the event and perhaps even the signal arrival time, are not known before the trigger is formed. Such tests, that appear to be real alerts, would constitute a more rigorous examination of alert handling and escalation procedures. Online tests may also be extended to the coordination of alerts with SNEWS, which may yield an additional host of benefits. Coordinated ``Fire Drill'' tests refer to issuing multiple CCSN alerts from different SNEWS-member neutrino experiments that are coincident with one another. This may be used to test SNEWS' ability to triangulate the position of a candidate in the sky, as well as ensure that astronomers are able to aim their instruments and properly identify a potential transient, if there is one. IceCube may also monitor SNEWS for an alert that is not coincident with an IceCube event, and test alert escalation and handling when no clear collective rate deviation is present in the detector. Knowing the proper procedures for handling and interpreting a supernova signal ensures that IceCube and SNEWS are prepared to issue a prompt notice of the detection and present accurate findings based on that detection.

\bibliographystyle{ICRC}
\bibliography{references}

\providecommand{\href}[2]{#2}\begingroup\raggedright\begin{thebibliography}{10}

\bibitem{Horiuchi:2017sku}
S.~Horiuchi and J.~P. Kneller
  \href{http://dx.doi.org/10.1088/1361-6471/aaa90a}{{\em J. Phys. G} {\bfseries
  45} no.~4, (2018) 043002}.

\bibitem{Nakamura:2016kkl}
K.~Nakamura {\em et~al.} \href{http://dx.doi.org/10.1093/mnras/stw1453}{{\em
  Mon. Not. Roy. Astron. Soc.} {\bfseries 461} no.~3, (2016) 3296--3313}.

\bibitem{Vissani:2010fg}
F.~Vissani, G.~Pagliaroli, and F.~Rossi-Torres. [Int. J. Mod.
  Phys.D20,1873(2011)].

\bibitem{Aartsen:2016}
{\bfseries IceCube} Collaboration, M.~Aartsen {\em et~al.} {\em JINST}
  {\bfseries 12} (2017) P03012.

\bibitem{Collaboration:2011ym}
{\bfseries IceCube} Collaboration, R.~Abbasi {\em et~al.}
  \href{http://dx.doi.org/10.1016/j.astropartphys.2012.01.004}{{\em Astropart.
  Phys.} {\bfseries 35} (2012) 615--624}.

\bibitem{Abbasi:2011ss}
{\bfseries IceCube} Collaboration, R.~Abbasi {\em et~al.}
  \href{http://dx.doi.org/10.1051/0004-6361/201117810e,
  10.1051/0004-6361/201117810}{{\em Astron. Astrophys.} {\bfseries 535} (2011)
  A109}.

\bibitem{Heereman:2015mbs}
D.~F. Heereman~von Zuydtwyck, {\em {HitSpooling}: {An Improvement for the
  Supernova Neutrino Detection System in IceCube}}.
\newblock PhD thesis, U. Brussels (main), 2015.

\bibitem{Kharusi:2020ovw}
{\bfseries SNEWS} Collaboration, S.~Al~Kharusi {\em et~al.}
  \href{http://dx.doi.org/10.1088/1367-2630/abde33}{{\em New J. Phys.}
  {\bfseries 23} no.~3, (2021) 031201}.

\bibitem{O_Connor_2011}
E.~O{\textquotesingle}Connor and C.~D. Ott
  \href{http://dx.doi.org/10.1088/0004-637x/730/2/70}{{\em The Astrophysical
  Journal} {\bfseries 730} no.~2, (Mar, 2011) 70}.

\bibitem{Piegsa:2009zz}
A.~Piegsa, ``{Supernova-Detektion mit dem IceCube-Neutrinoteleskop},'' other
  thesis, 9, 2009.

\bibitem{Cross:2019jpb}
{\bfseries IceCube} Collaboration, R.~Cross, A.~Fritz, and S.~Griswold
  \href{http://dx.doi.org/10.22323/1.358.0889}{{\em PoS} {\bfseries ICRC2019}
  (2020) 889}.

\bibitem{Baum:2013}
{\bfseries IceCube} Collaboration, V.~Baum, D.~Heereman, and R.~Bruijn {\em
  Braz. J. Phys.} {\bfseries 44} no.~5, (2014) 0444.

\bibitem{Baum:2015drl}
{\bfseries IceCube} Collaboration, V.~Baum, B.~Eberhardt, A.~Fritz,
  D.~Heereman, and B.~Riedel \href{http://dx.doi.org/10.22323/1.236.1096}{{\em
  PoS} {\bfseries ICRC2015} (2016) 1096}.

\bibitem{Baum:2017rty}
V.~Baum, \href{http://dx.doi.org/10.25358/openscience-1372}{{\em {Search for
  Low Energetic Neutrino Signals from Galactic Supernovae and Collisionally
  Heated Gamma-Ray Bursts with the IceCube Neutrino Observatory}}}.
\newblock PhD thesis, Mainz U., 2017.

\bibitem{ASTERIA:2019}
``{ASTERIA: A Supernova TEst Routine for IceCube Analysis},'' 2019.
\newblock \url{https://github.com/IceCubeOpenSource/ASTERIA/}.

\bibitem{Nakazato:2012qf}
K.~Nakazato {\em et~al.}
  \href{http://dx.doi.org/10.1088/0067-0049/205/1/2}{{\em Astrophys. J. Suppl.}
  {\bfseries 205} (2013) 2}.

\end{thebibliography}\endgroup


\providecommand{\href}[2]{#2}\begingroup\raggedright\endgroup

\clearpage
\section*{Full Author List: IceCube Collaboration}

% \noindent \textbf{Note comment afterwards:} Collaborations have the possibility to provide an authors list in xml format which will be used while generating the DOI entries making the full authors list searchable in databases like Inspire HEP. For instructions please go to icrc2021.desy.de/proceedings or contact us under icrc2021proc@desy.de.\\

% \scriptsize
% \noindent
% first.author$^1$, 
% second.author$^2$, 
% third.author$^3$ % .... more names
% and 
% last.author$^{n}$ \\

% \noindent
% $^1$first.affiliation.
% $^2$second.affiliation. % .... more affiliation
% $^{m}$last.affiliation.

\scriptsize
\noindent
R. Abbasi$^{17}$,
M. Ackermann$^{59}$,
J. Adams$^{18}$,
J. A. Aguilar$^{12}$,
M. Ahlers$^{22}$,
M. Ahrens$^{50}$,
C. Alispach$^{28}$,
A. A. Alves Jr.$^{31}$,
N. M. Amin$^{42}$,
R. An$^{14}$
% ...
\\

\noindent
% ...
$^{12}$ Universit{\'e} Libre de Bruxelles, Science Faculty CP230, B-1050 Brussels, Belgium \\
% ...
$^{14}$ Department of Physics and Laboratory for Particle Physics and Cosmology, Harvard University, Cambridge, MA 02138, USA \\
% ...
$^{17}$ Department of Physics, Loyola University Chicago, Chicago, IL 60660, USA \\
$^{18}$ Dept. of Physics and Astronomy, University of Canterbury, Private Bag 4800, Christchurch, New Zealand \\
% ...
$^{22}$ Niels Bohr Institute, University of Copenhagen, DK-2100 Copenhagen, Denmark \\
% ...
$^{28}$ D{\'e}partement de physique nucl{\'e}aire et corpusculaire, Universit{\'e} de Gen{\`e}ve, CH-1211 Gen{\`e}ve, Switzerland \\
% ...
$^{31}$ Karlsruhe Institute of Technology, Institute for Astroparticle Physics, D-76021 Karlsruhe, Germany  \\
% ...
$^{42}$ Bartol Research Institute and Dept. of Physics and Astronomy, University of Delaware, Newark, DE 19716, USA \\
% ...
$^{50}$ Oskar Klein Centre and Dept. of Physics, Stockholm University, SE-10691 Stockholm, Sweden \\
% ...
$^{59}$ DESY, D-15738 Zeuthen, Germany \\
% ...

\subsection*{Acknowledgements}

\noindent
USA {\textendash} U.S. National Science Foundation-Office of Polar Programs,
U.S. National Science Foundation-Physics Division,
U.S. National Science Foundation-EPSCoR,
Wisconsin Alumni Research Foundation,
Center for High Throughput Computing (CHTC) at the University of Wisconsin{\textendash}Madison,
Open Science Grid (OSG),
Extreme Science and Engineering Discovery Environment (XSEDE),
Frontera computing project at the Texas Advanced Computing Center,
U.S. Department of Energy-National Energy Research Scientific Computing Center,
Particle astrophysics research computing center at the University of Maryland,
Institute for Cyber-Enabled Research at Michigan State University,
and Astroparticle physics computational facility at Marquette University;
Belgium {\textendash} Funds for Scientific Research (FRS-FNRS and FWO),
FWO Odysseus and Big Science programmes,
and Belgian Federal Science Policy Office (Belspo);
Germany {\textendash} Bundesministerium f{\"u}r Bildung und Forschung (BMBF),
Deutsche Forschungsgemeinschaft (DFG),
Helmholtz Alliance for Astroparticle Physics (HAP),
Initiative and Networking Fund of the Helmholtz Association,
Deutsches Elektronen Synchrotron (DESY),
and High Performance Computing cluster of the RWTH Aachen;
Sweden {\textendash} Swedish Research Council,
Swedish Polar Research Secretariat,
Swedish National Infrastructure for Computing (SNIC),
and Knut and Alice Wallenberg Foundation;
Australia {\textendash} Australian Research Council;
Canada {\textendash} Natural Sciences and Engineering Research Council of Canada,
Calcul Qu{\'e}bec, Compute Ontario, Canada Foundation for Innovation, WestGrid, and Compute Canada;
Denmark {\textendash} Villum Fonden and Carlsberg Foundation;
New Zealand {\textendash} Marsden Fund;
Japan {\textendash} Japan Society for Promotion of Science (JSPS)
and Institute for Global Prominent Research (IGPR) of Chiba University;
Korea {\textendash} National Research Foundation of Korea (NRF);
Switzerland {\textendash} Swiss National Science Foundation (SNSF);
United Kingdom {\textendash} Department of Physics, University of Oxford.

\end{document}